\documentclass[aps,pra,superscriptaddress,amsmath,amssymb,preprintnumbers,twocolumn,notitlepage]{revtex4}
\usepackage{qcircuit}
\usepackage{amsmath,bm}
\usepackage[dvips]{graphicx}
\usepackage{amsmath,amssymb,amsthm,mathrsfs,amsfonts,dsfont}
\usepackage{braket}
\usepackage{bm}
\usepackage{enumerate}
\usepackage{color}
\usepackage{graphicx}
\usepackage{algorithm}
\usepackage{algorithmic}
\usepackage{ulem}
\usepackage[justification=RaggedRight]{caption}
\usepackage[subrefformat=parens]{subcaption}
\usepackage{here}

\begin{document}


\title{Analysis of the shortest vector problems with the quantum annealing to search the excited states}

\author{Katsuki Ura \footnote{These authors equally contributed to this paper.}
}
\affiliation{Department  of  Physics, Tokyo  University  of  Science,  Shinjuku,  Tokyo  162-8601,  Japan.}
\affiliation{Research Center for Emerging Computing Technologie, National Institute of Advanced Industrial Science and Technology (AIST), Umezono1-1-1, Tsukuba, Ibaraki 305-8568, Japan.}

\author{Takashi Imoto$^*$
}
\affiliation{Research Center for Emerging Computing Technologie, National Institute of Advanced Industrial Science and Technology (AIST), Umezono1-1-1, Tsukuba, Ibaraki 305-8568, Japan.}

\author{Tetsuro Nikuni}\email{nikuni@rs.tus.ac.jp}
\affiliation{Department  of  Physics,  Faculty  of  Science  Division  I,Tokyo  University  of  Science,  Shinjuku,  Tokyo  162-8601,  Japan.}

\author{Shiro Kawabata}\email{s-kawabata@aist.go.jp}
\affiliation{Research Center for Emerging Computing Technologie, National Institute of Advanced Industrial Science and Technology (AIST), Umezono1-1-1, Tsukuba, Ibaraki 305-8568, Japan.}
\affiliation{NEC-AIST Quantum Technology Cooperative Research Laboratory, National Institute of Advanced Industrial Science and Technology (AIST), Tsukuba, Ibaraki 305-8568, Japan}

\author{Yuichiro Matsuzaki}\email{matsuzaki.yuichiro@aist.go.jp}
\affiliation{Research Center for Emerging Computing Technologie, National Institute of Advanced Industrial Science and Technology (AIST), Umezono1-1-1, Tsukuba, Ibaraki 305-8568, Japan.}
\affiliation{NEC-AIST Quantum Technology Cooperative Research Laboratory, National Institute of Advanced Industrial Science and Technology (AIST), Tsukuba, Ibaraki 305-8568, Japan}


\begin{abstract}
The shortest vector problem (SVP) is one of the lattice problems and is mathematical basis for the lattice-based cryptography, which is expected to be  post-quantum cryptography. The SVP can be mapped onto the Ising problem, which in principle can be solved by quantum annealing (QA). However, one issue in solving the SVP using QA is that the solution of the SVP corresponds to the first excited state of the problem Hamiltonian. Therefore, QA, which searches for ground states, cannot provide a solution with high probability. In this paper, we propose to adopt an excited-state search of the QA to solve the shortest vector problem. We numerically show that the excited-state search provides a solution with a higher probability than the ground-state search.
\end{abstract}

\maketitle

\section{Introduction}
Quantum annealing (QA) \cite{apolloni1989quantum,finnila1994quantum,morita2008mathematical,hauke2020perspectives} is one of the methods to solve combinatorial optimization problems \cite{kadowaki1998quantum,lucas2014ising}.
It is known that the combinatorial optimization problem \cite{schrijver2005history} can be mapped as the problem of finding the ground state of the Ising Hamiltonian \cite{lechner2015quantum}.
QA is used to find the ground state of the Ising Hamiltonian.
In QA, the Hamiltonian is time-dependent; we slowly change the Hamiltonian from the independent spin model with the transverse fields (called the driver Hamiltonian) to the target Ising Hamiltonian (called a problem Hamiltonian).
As long as an adiabatic condition holds \cite{farhi2000quantum,aharonov2008adiabatic,farhi2001quantum,albash2018adiabatic,childs2001robustness}, we can obtain a ground state of the problem Hamiltonian when the initial state is a ground state of the driver Hamiltonian.

D-Wave Systems Inc. has developed a quantum hardware to perform QA using thousands of superconducting flux qubits \cite{mcgeoch2019practical,johnson2011quantum,boixo2014evidence,boixo2013experimental}.
Several other quantum hardware for QA have been proposed and developed\cite{maezawa2019toward,saida2021experimental,saida2022factorization}.
Previous researches mainly focused on the ground-state search for QA.

More recently, there have been studies of excited state
searches in which the excited state of the driver Hamiltonian is
selected as the initial state \cite{seki2021excited,teplukhin2019calculation,teplukhin2020electronic,chen2020demonstration}.
A crucial point of the excited state search is that we need to use non-uniform transverse magnetic fields in the driver Hamiltonian to resolve the degeneracy of the excited state of the driver Hamiltonian.
This procedure allows us to prepare a non-degenerate excited state of the driver Hamiltonian when we start QA.
By changing the Hamiltonian from the driver one to the problem one, we obtain the excited state of the problem Hamiltonian as long as the adiabatic condition is satisfied.
The excited-state search in QA is useful in quantum chemistry\cite{serrano2005quantum}; for example, it is essential to know the photochemical properties of molecules, which requires information not only on the ground state but also on the excited state.

Post-quantum cryptography
has attracted much attention
from many researchers.
RSA (Rivest–Shamir–Adleman) is a widely used public key cryptography with the security based on the difficulty of the prime factorization\cite{rivest1978method}. 
Once the fault-tolerant quantum computer is developed, RSA cryptography can be efficiently decrypted by Shor's algorithm \cite{shor1994algorithms}.
Therefore, research on post-quantum cryptography, which is difficult to solve even with a gate-type quantum computer, is underway.
Lattice-based cryptography (LBC) \cite{ajtai1997public} is one of the candidates for post-quantum cryptography \cite{micciancio2009lattice}.

One of the key mathematical problems in LBC is the shortest vector problem (SVP), which is the problem of finding the shortest non-zero vector in a given lattice.
There are
two approaches 
to solving lattice problems.
The first approach is to chose input vectors from a distribution on a lattice and iteratively combine the vectors so that output should be probabilistically generated as solutions
\cite{ajtai2001sieve, laarhoven2015sieving, micciancio2010faster, ducas2018shortest}.
The second approach is to
enumerate all vectors in a specific sphere centered at the origin. There is a guarantee that the solution is contained if it is carefully chosen\cite{pohst1981computation, gama2010lattice}.
Although these are classical algorithms, it is known that the SVP can be solved using a gate-type quantum computer. A quantum tree algorithm (based on Grover's algorithm) can solve the SVP \cite{laarhoven2013solving}. However, it still takes an exponentially longer time to solve larger problems.

Recently, a method using quantum annealing was proposed to search for solutions of the SVP. More specifically,
Joseph et al. proposed a heuristic method for finding the solution to the shortest vector problem using ground-state search for QA \cite{joseph2021two}.
The SVP can be mapped onto an Ising Hamiltonian with integer spins, and the first excited state of the Hamiltonian corresponds to the solution.
In their method, after the ground state of the driver Hamiltonian is prepared, the Hamiltonian is changed from the driver Hamiltonian to the problem Hamiltonian over time.
The goal is to obtain the desired first excited state of the problem Hamiltonian via a non-adiabatic transition from the ground state to the first excited state.
However, there is no known method to find a suitable schedule to change the Hamiltonian for obtaining the first excited state in their approach.
If the Hamiltonian is changed slowly, the ground state is obtained.
On the other hand, a rapid change of the Hamiltonian would induce non-adiabatic transitions to not only the first excited state but also to other excited states.

In this paper, we propose to use the excited-state search with QA to find a solution to the SVP.
We adopt the inhomogeneous transverse fields with integer spins as the driver Hamiltonian so that we can prepare the non-degenerate first excited state of the driver Hamiltonian. By changing the Hamiltonian from the driver Hamiltonian to the problem Hamiltonian, we can obtain with finite probability the first excited state of the problem Hamiltonian, which is the solution of the SVP. By increasing the annealing time, the dynamics become more adiabatic, and
the success probability should increase as long as the decoherence is negligible. 

We also show that the first excited state of the driver Hamiltonian with integer spins is an entangled state, which is experimentally challenging to prepare.
We show that it is still possible to obtain the first excited state with a high probability in our method by using a specific separable state as the initial state \cite{kitagawa1993squeezed,radcliffe1971some,arecchi1972atomic}.
Moreover, we compare our method based on the excited-state search with the previous approach based on the ground-state search. We show that our method provides higher success probabilities for most of the parameters.

\section{Quantum Annealing (QA)}
\subsection{ground-state search}
In this subsection, we describe QA for the ground-state search.
In QA, quantum fluctuations are used to find a ground state of a given Ising Hamiltonian.
The total Hamiltonian for QA is given as follows:
\begin{eqnarray}
H(t)=\biggl(1-\frac{t}{T}\biggr)H_\mathrm{D}+\biggl(\frac{t}{T}\biggr)H_\mathrm{P},
\end{eqnarray}
\begin{eqnarray}
 H_\mathrm{D}=-b_x\sum^{N}_{i=1} \sigma_x^{(i)},
\end{eqnarray}
\begin{eqnarray}
H_\mathrm{P}=\sum^{N}_{i=1} h_{i} \sigma_z^{(i)}
+ \sum^{N}_{i=1} J_{i,j} \sigma_z^{(i)}\sigma_z^{(j)},
\end{eqnarray}
where $H_D$ is the driver Hamiltonian to induce quantum fluctuations, $H_P$ is the problem Hamiltonian whose ground state corresponds to the solution of the combinational optimization problem, $b_x$ denotes the strength of the transverse magnetic field, $h_i$ denotes the longitudinal magnetic field, and $J_{i,j}$ denotes the coupling constant of the Ising interaction.
Also, 
$\sigma _x$ and $\sigma _z$ 
denote the Pauli matrices.

We prepare an initial state as the ground state of the driver Hamiltonian, and let this state evolve by the time-dependent Hamiltonian $H(t)$. As long as the dynamics is adiabatic, we can obtain the ground state of the problem Hamiltonian at $t=T$.
On the other hand, if the dynamics is not slow enough to satisfy the adiabatic condition, non-adiabatic transitions occur, and there will be a finite population in the excited states.

\subsection{Excited state search}
In this section, we describe QA for the excited state search\cite{seki2021excited,teplukhin2019calculation,teplukhin2020electronic,chen2020demonstration}.
We consider the following Hamiltonian
\begin{eqnarray}
H(t)=\biggl(1-\frac{t}{T}\biggr)H_\mathrm{D}^{{(\mathrm{nu})}}+\biggl(\frac{t}{T}\biggr)T_P
\end{eqnarray}
\begin{eqnarray}
H_\mathrm{D}^{(\mathrm{nu)}}=-\sum^{N}_{i=1} b_x^{(i)}\sigma_x^{(i)}
\end{eqnarray}
\begin{eqnarray}
H_\mathrm{P}=\sum^{N}_{i=1} h_{i} \sigma_z(i)
+
\sum^{N}_{i=1} J_{i,j} \sigma_z(i) \sigma_z(j)
\end{eqnarray}
where $b_x^{(i)}$ is the amplitude of the transverse magnetic field at site $i$.
This spatially non-uniform transverse magnetic field can resolve the degeneracy of the first excited state of the driver Hamiltonian.
First, we prepare the first excited state of $H_D$.
Second, we let the system evolve according to the time-dependent Hamiltonian from $t=0$ to $t=T$. After these steps, we can obtain the first excited state of the problem Hamiltonian, as long as the adiabatic condition is satisfied.

\section{The shortest vector problem (SVP)}
We review the shortest vector problem (SVP), which is the mathematical basis for post-quantum cryptography. We consider a set of lattice vectors as defined below:
\begin{eqnarray}
\boldsymbol{L}=\{\sum^{N}_{i=1} {x_{i}\Vec{\boldsymbol{b}}_{i}}\}=\{{\boldsymbol{B}\cdot\boldsymbol{x}:\boldsymbol{x}\in \mathbb{Z}^{N}}\}
\end{eqnarray}
where $\boldsymbol{x}=\{ x_{i} \}_{i=1}^N\in \mathbb{Z}^{N}$ is a set of integers representing the coefficients of the lattice basis vectors, $\boldsymbol{x}$
is a vector of the coefficients, 
$\{ \boldsymbol{b}_{i} \}^N_{i=1}$
is a set of linearly independent vectors, and $\boldsymbol{B}=\{\boldsymbol{b}_{1}, \boldsymbol{b}_{2}, \cdots \boldsymbol{b}_{N} \}$ is the lattice basis matrices. Each vector on the lattice is expressed as follows.
\begin{eqnarray}
\boldsymbol{v}=\boldsymbol{B}\cdot\boldsymbol{x}
=x_{1}\boldsymbol{b}_{1}+\cdots +
x_{N}\boldsymbol{b}_{N}\in \boldsymbol{L}
\end{eqnarray}
The SVP aims to find a non-zero vector with the smallest norm on this lattice. 

\section{Mapping of SVP}
Let us explain how to map the SVP onto the Ising Hamiltonian \cite{joseph2021two}. The norm of the vector $\boldsymbol{v}$ on the lattice can be written as 
\begin{eqnarray}
&&||\boldsymbol{v}||^{2}=\sum^{N}_{i,j=1} x_{i}x_{j}\boldsymbol{b}_{i}\cdot\boldsymbol{b}_{j}\nonumber \\
&=&
\sum^{N}_{i,j=1} x_{i}x_{j} \boldsymbol{G}_{i,j}
\end{eqnarray}
where $\boldsymbol{G}_{i,j}=\boldsymbol{b}_{i}\cdot\boldsymbol{b}_{j}$ is the element of the Gram matrix of the lattice basis vectors. 
We consider the search for a solution of the SVP in the range  $-k \leq x_i\leq k$ . 
Let us consider $2kN$ qubits, and
the Hamiltonian corresponding to the norm can be written as 
\begin{eqnarray}
\hat{H}^{(\rm{SVP})}_{\rm{p}}=J\sum^{N}_{i,j=1} \boldsymbol{G}_{i,j}\hat{Q}^{(i)}\hat{Q}^{(j)}\label{hsvp}
,
\end{eqnarray}
where $\hat{Q}^{(i)}$ is a diagonal matrix defined as
$\hat{Q}^{(i)}=\sum^{2k}_{p=1}\hat{\sigma}^{(p,i)}_{z}/2$ ($i=1,2,\cdots, N$),
$J$ denotes a constant factor with a unit of energy,
and $\sigma_z$ denotes a Pauli matrix. 
Throughout of this paper, by setting $J=1$, the time and energy are normalized by this value. 
To save the computational resources, we consider a subspace spanned by Dicke basis. Then,
the eigenvalue of the operator
$\hat{Q}^{(i)}$ corresponds to the coefficient of the $N$ lattice basis vectors, which takes the integer value
in the range of $-k \leq x_i\leq k$. 
The ground state of the Hamiltonian (10) corresponds to the zero vector. Therefore, the first excited state is the solution of the SVP.

\section{Solving SVP using a ground state search using adiabatitic transition with QA}
We briefly explain the previous study \cite{joseph2021two} on finding the solution of the SVP using the ground state search with QA. The driver Hamiltonian is described as 
\begin{eqnarray}
\hat{H}^{\rm{SVP}}_{D}=\sum _{i=1}^N B_{x}
\sum^{2k}_{p=1}\hat{\sigma}^{(p,i)}_{x}
,
\end{eqnarray}
where $B_x$ is the strength of the transverse magnetic field and $\hat{\sigma_x}$ denotes the Pauli operator. We adopt $H_P^{(SVP)}$ as the problem Hamiltonian. The total Hamiltonian is given as
\begin{eqnarray}
\hat{H(t)}=\biggl(1-\frac{t}{T}\biggr)\hat{H_D}^{(SVP)}+\biggl(\frac{t}{T}\biggr)\hat{H_P}^{(SVP)}
,
\end{eqnarray}
where $T$ is the annealing time. The QA was originally proposed to find the ground state of the problem Hamiltonian with the adiabatic dynamics. After preparing the ground state of the driver Hamiltonian, we evolve the system according to the total Hamiltonian from $t=0$ to $t=T$. As long as the dynamics is adiabatic, the ground state of the problem Hamiltonian is obtained as a final state. However, since the first excited state of the problem Hamiltonian is the solution of the SVP, we cannot obtain the solution with high probability by using the ground-state search. In the previous approach, non-adiabatic transitions are utilized to excite the system. If one could find suitable scheduling, we may obtain the first excited state with high probability. However, finding an optimal annealing time is
not straightforward
as long as the ground-state search is used.


\section{Solving SVP using excited state search with QA}
In this section, we propose a method for finding a solution to the SVP using the excited-state search with QA.

\subsection{Preparing the first excited state as the initial state}
For the excited-state search, the driver Hamiltonian is given by 
\begin{eqnarray}
\hat{H}^{\rm{(SVPE)}}_{D}=\sum _{i=1}^N B_x^{(i)}
\sum^{2k}_{p=1}\hat{\sigma}_x^{(p,i)}
,
\end{eqnarray}
where $\{{B_x}^{(i)}\}_{i=1}^N$ represents the strength of the non-uniform transverse magnetic field. 
We set $b_x^{(1)} < b_x^{(2)}=\cdots =b_x^{(N)}$, to resolve the degeneracy of the first excited state of the driver Hamiltonian.
On the other hand, we adopt $H_{P}^{\left(\mathrm{SVP}\right)}$ in Eq.(10) as the problem Hamiltonian. The total Hamiltonian is $H=(1-\frac{t}{T})H_D^{\left(\mathrm{SVPE}\right)}+\frac{t}{T}H_P^{\left(\mathrm{SVP}\right)}$. After we prepare the first excited state of $H_D^{\left(\mathrm{SVPE}\right)}$ as the initial state, we let the system evolve according to $H$ from $t=0$ to $t=T$.
The first excited state of the driver Hamiltonian
is described as 
\begin{eqnarray}
|W\rangle _{1-2k} \bigotimes _{j=2k+1}^{2Nk}|-\rangle _j
,
\end{eqnarray}
where 
$\bigotimes _{j=l}^{m}|-\rangle _j=|-\rangle _{l}|-\rangle _{l+1}\cdots |-\rangle_{m}$ denotes a separable state, and
$|W\rangle _{1-2k}$ is the entangled state given by.
\begin{eqnarray}
|W\rangle _{1-2k}=\frac{1}{\sqrt{2k}}\sum _{p=1}^{2k} \hat{\sigma}^{(p,1)}_z \bigotimes _{j=1}^{2k}|-\rangle _j.
\end{eqnarray}
Unlike the previous approach of Ref. [34], we change the Hamiltonian in an adiabarically so that we obtain the first excited state of $H_\mathrm{p}^{\left(\mathrm{SVP}\right)}$, which is the solution of the SVP. The adiabatic theorem guarantees that we can obtain the solution with a high probability by taking a sufficiently long time.


\subsection{Using spin coherent state for initial state}
The aforementioned excited state search requires quantum annealing to start from an initial state that contains an entanglement. However, preparing an entangled initial state in the actual QA device is challenging.
In actual experiments, it is desirable to use a separable initial state.
Therefore we consider using the spin-coherent (SC) state as the initial state.
The SC state is described as follows:
\begin{eqnarray}
\ket{\phi}_{1-2k}&=& \bigotimes _{j=1}^{2k} 
(\sqrt{\epsilon}|+\rangle _j + \sqrt{1-\epsilon}|-\rangle _j 
)
,
\end{eqnarray}
where we set
$\epsilon=\frac{1}{2k-2}$.
The inner product of $\ket{W_{1-2k}}$ and $\ket{\phi_{1-2k}}$ is calculated as
\begin{eqnarray}
{}_{1-2k}\braket{W|\phi}_{1-2k} =\sqrt{k\epsilon(1-\epsilon)^{2k-2}}\\
\fallingdotseq\sqrt{\frac{k}{2k-2}e^{-1}}\\
=\sqrt{\frac{1}{2e}}\approx 0.43\\
(k\gg1)
\end{eqnarray}
This means that the SC state contains the first excited state $\ket{W}_{1-2k}$ with a reasonably high probability.
Therefore, we propose to use the SC state as an initial state for the excited state search to solve the SVP.
We will perform numerical simulations to quantify the performance of the search with the SC state.

\section{Numerical calculation}
In this section, we show numerical results to compare the performance of our scheme with that of the previous scheme [1]. We consider the SVP with $N=2 , i.e.,$ the two-dimensional lattice. The two vectors $\boldsymbol{b_1}$ and $\boldsymbol{b_2}$ are given in the problem. We characterize these vectors
by their norms $\{b_j\}_{j=1}^2$ and the angle $\theta$ between them. By solving the time-dependent Schrödinger equation, we obtain the state after QA. 
We define the failure probability as the probability that the measurement result (in computational basis)
for the state after QA gives an incorrect answer for the SVP. 
This means that the success probability is defined as a population of the first excited state of the problem Hamiltonian after QA.
We calculate the failure probabilities for the ground-state and excited-state search, respectively. 
For the ground-state search, we optimize $B_x$ to minimize the failure probability. On the other hand, for the excited-state search, we optimize $B_x^{(1)}$ to minimize the failure probability while fixing $B_x^{(1)}/B_x^{(2)}$ at a specific value. We set $k=2$, and thus the spin quantum number is 2 in our numerical simulation. 
We fix the norm of the vectors to $b_1=b_2$, and change the values of $\theta$ such as $\frac{\pi}{18},\frac{\pi}{9},\frac{\pi}{6}$. 
In Fig. 1, we plot the failure probabilities for the ground state and excited-state search against the annealing time $T$, respectively.
These results show that the failure probability for the excited-state search is smaller than that for the ground state search. In the excited state search, the failure probability with $\theta=\frac{\pi}{18}$ is larger than those with $\theta=\frac{\pi}{9},\frac{\pi}{6}$. 
This may correspond to the fact that in the SVP, the problem becomes more difficult as the angle $\theta$ becomes smaller. However, the primary advantage of our method is that, even if the problem becomes more difficult, the failure probability can be sufficiently small by taking a reasonably long annealing time $T$, which is guaranteed by the adiabatic theorem.

\begin{figure}[H]
\centering
\includegraphics[width=80mm]{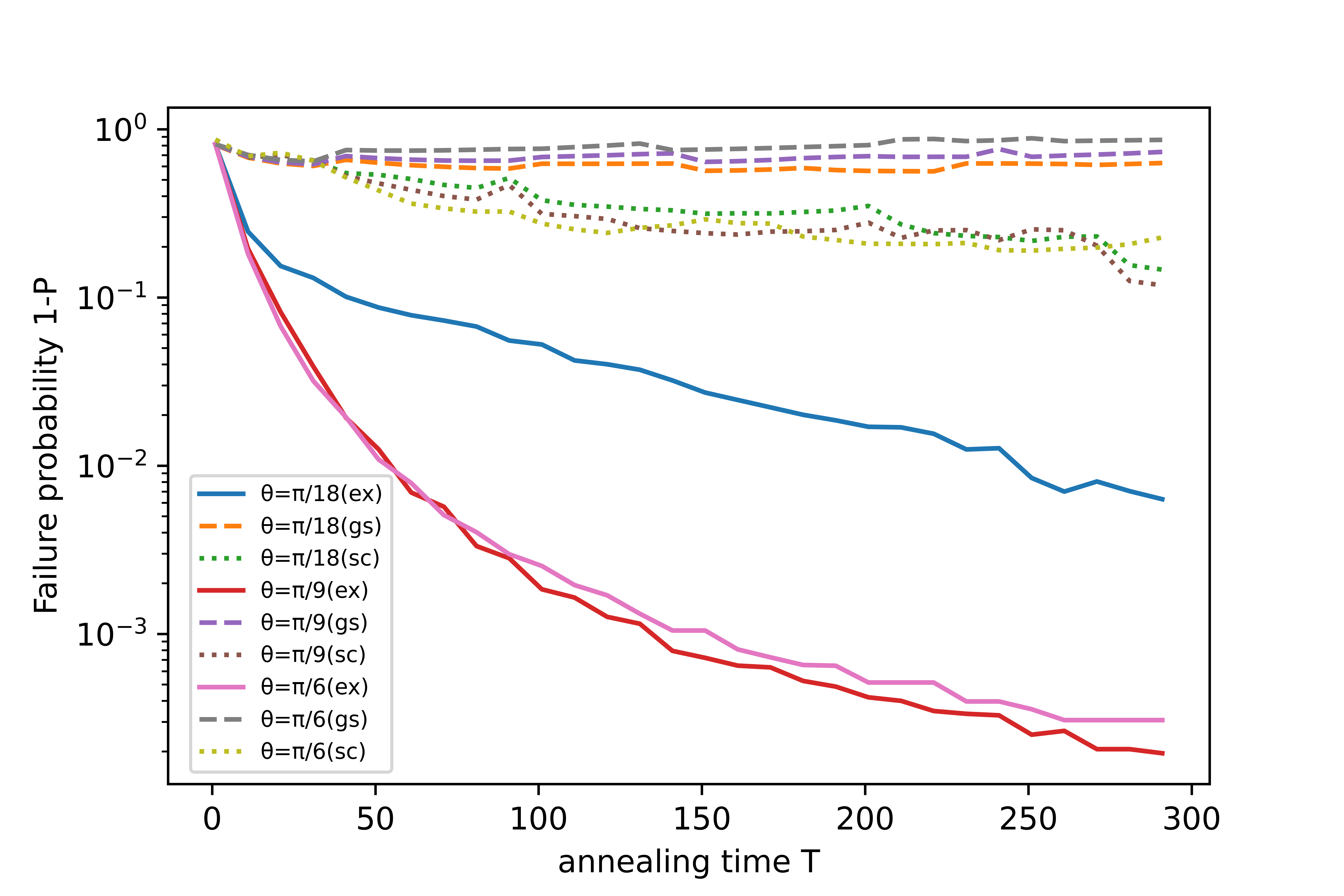}
\caption{Plot of the failure probability of QA for the excited state search, the search with a spin coherent state, the ground state search against an annaling time $T$. 
For most of the parameters, the excited state search and the search with the spin coherent state provides a smaller failure probability than the ground state search. We use $k=2$ and $N=2$ and set $b_1=b_2=1$.
We fix
 $B_x^{(1)}/B_x^{(2)}=1/2$ for the excited state search and the search with a spin coherent state, while we fix $B_x^{(1)}/B_x^{(2)}=1$ for the ground state search.
 Moreover, we optimize the values of $B_x^{(1)}$ to minimize the failure probability. 
}\label{thetavsp}
\end{figure}

The failure probability of the search with the SC state is larger than that using the first excited state.
This is because the SC state includes states other than the first excited state.
On the other hand, the search with the SC state
provides a smaller
failure probability 
the ground state search when the annealing time $T$ is large, and this shows the practicality of our scheme.

Next, we consider the case where the ratio of the norms of the two vectors is fixed at either 1:1 or 1:2. We then plot how the failure probability changes when the angle between the vectors is changed. In each case, the amplitude of the transverse magnetic field is optimized to minimize the failure probability.
Figure 2 shows that the failure probability is always lower for the excited state search using the first excited state than that for the ground state search.
The search with SC state also shows a smaller failure probability than the ground state search, except for a few exceptional points (where the ratio of vectors is 1:2 and the angle is around $\pi/2$).
The reason why the failure probability of the excited state search is larger at an angle of $\pi/2$ when the ratio of vectors is 1:2 is due to the existence of the symmetry in the Hamiltonian, which causes energy-level crossing in quantum annealing.
This point is explained in detail in the Appendix.

\begin{figure}[H]
\centering
\includegraphics[width=80mm]{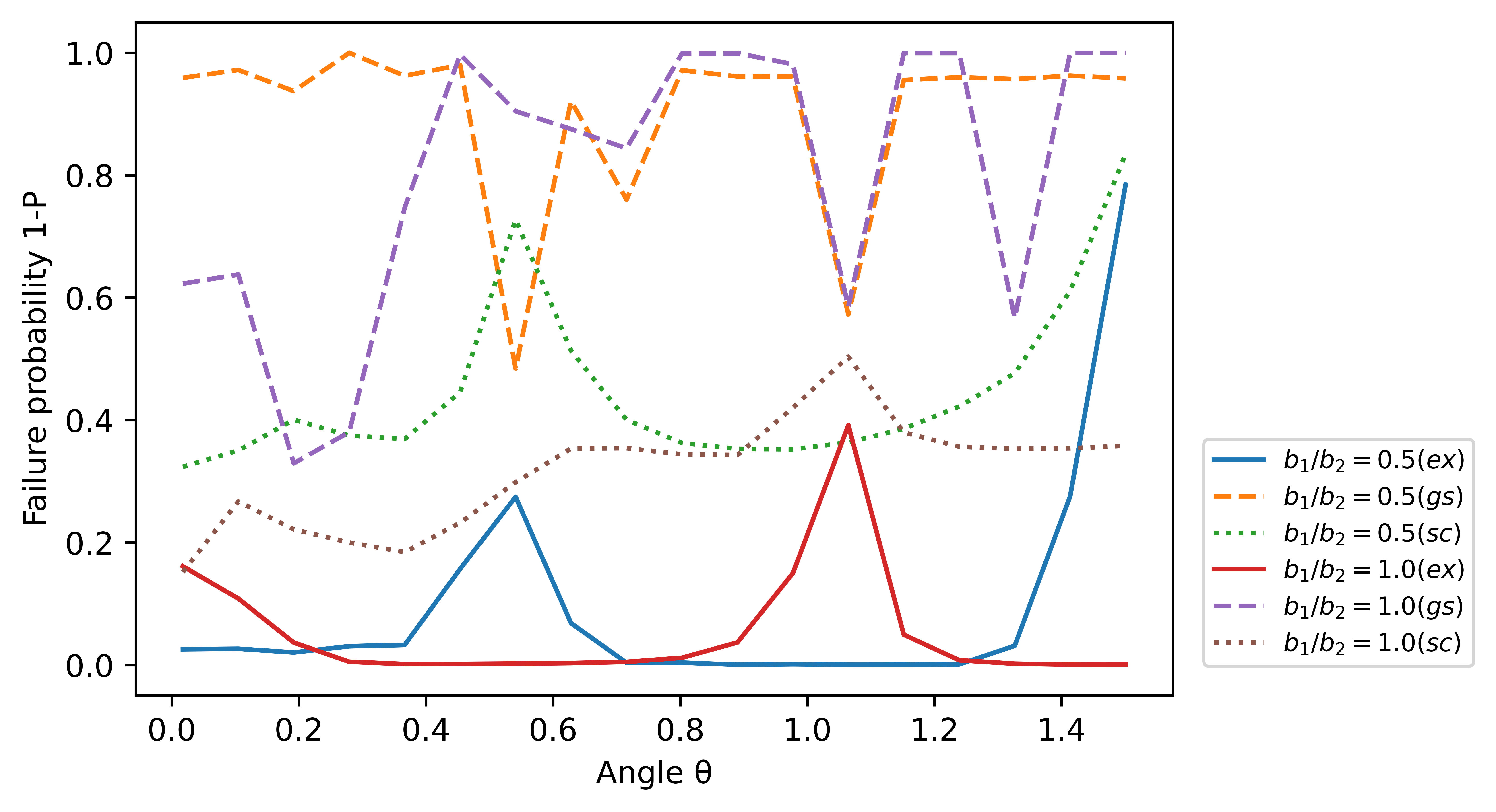}
\caption{
Plot of the failure probability of QA 
We plot the results for the excited state search (ex), the search with
the spin coherent state (sc), the ground state search (gs), respectively, against the angle $\theta$ where we fix the value of $b_1/b_2$. This graph shows that there is a specific angle at which the probability of failure increases.
We use $k=2$ and $N=2$.
We fix
 $B_x^{(1)}/B_x^{(2)}=1/2$ for the excited state search and the search with a spin coherent state, while we fix $B_x^{(1)}/B_x^{(2)}=1$ for the ground state search.
 Moreover, we optimize the values of $B_x^{(1)}$ to minimize the failure probability. Also, we fix $T=100$ for the excited state search and the search with a spin coherent state
 while we optimize $T$ for the ground state search.
}\label{thetavsp}
\end{figure}

\begin{figure}[H]
\centering
\includegraphics[width=80mm]{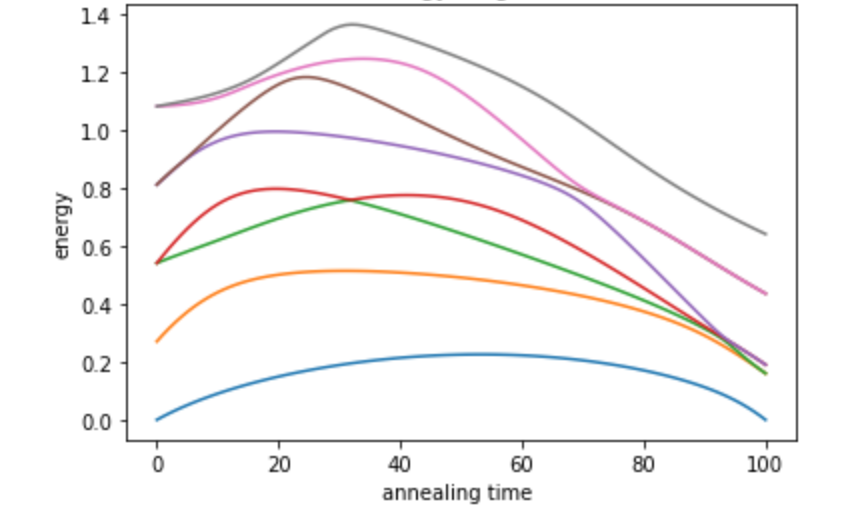}
\caption{We plot the instantaneous eigenenergy of the total Hamiltonian during QA or the vector ratio is 1:2 and the angle
is $\pi/6$.
This graph shows that the energy gaps among the first, second, third, and fourth excited states is very small.
Except the vector ration and the angle these, we use the same parameters as used in
the Fig. \ref{thetavsp}.
}\label{thetavsptwo}
\end{figure}

In the case of the excited state search, when the vector ratio is 1:2, the failure probability becomes large around the angle of $\pi/6$.
This is due to the fact that, the energies of the the lowest four excited states of $H_p^{(SVP)}$ are close to each other, as shown in Fig. \ref{thetavsptwo}.
The small energy gap causes non-adiabatic transitions from the first excited state to the other excited states, which increases the failure probability.
Also, when the vector ratio is 1:1, the failure probability is larger around the angle of $\pi/3$.
This is due to the fact that the first excited state of $H^{(\rm{SVP})}_{\rm{p}}$ is 6-fold degenerate 
when the vector ratio is 1:1 and the angle is $\pi/3$.
In this case, $(x_{1},x_{2})=(1,0),\ (0,1)\ (1,-1),\ (-1,0),\ (0,-1),\ (-1,1)$ provide the shortest vector.
Therefore, the energy gap between the first excited state and the other excited states becomes smaller at angles around $\pi/3$, resulting in more non-adiabatic transitions.

\section{Possible physical realization }
To implement the excited state search with QA, it is crucial to use a long-lived qubit, because otherwise the excited state would decay into the ground state by the energy relaxation. There was a theoretical proposal to perform QA with capacitively-shunted flux qubits\cite{matsuzaki2020quantum}, which has a long coherence time such as tens of microseconds. Therefore, we expect that our proposed protocol can be demonstrated by the capacitively-shunted flux qubits. 

\section{Conclusion}
In conclusion, we propose to use the excited-state search with the QA to solve the SVP. 
Importantly, the solution of the SVP is not the ground state but the first excited state of the problem Hamiltonian. So,
unlike the previous approach of solving the SVP by ground-state search using QA, the adiabatic theorem guarantees that our scheme can obtain a solution in our approach
if we take a sufficiently long time. 
Our numerical simulations reveal that our scheme provides a smaller failure probability than the previous scheme. Our results show the potential of our scheme to solve the SVP by using a quantum annealer. 
However, to satisfy the adiabatic condition with our methods, it may take an exponentially long annealing time with the system size if the energy gap is exponentially small, depending on the problems.
It is essential to classify which problems are difficult to solve due to such an exponentially small energy gap for QA. We leave this point for future work.
\begin{acknowledgments}
K. U. and T. I.  contributed to this
work equally.
This work was supported by MEXT’s Leading Initiative for Excellent Young Researchers, JST PRESTO (Grant No. JPMJPR1919), Japan. This paper is partly based on the results obtained from the project, JPNP16007, commissioned by the New Energy and Industrial Technology Development Organization (NEDO), Japan.
\end{acknowledgments}


\appendix
\section{}
In this appendix,
we explain how the symmetry appears in
the excited state search at the angle of $\pi/2$ when the ratio of vectors is 1:2, which causes nonadiabatic transitions in quantum annealing.

Let us define the parity operator $\hat{P}_i=e^{-i\pi\hat{S}^{(i)}_x }$ for spin $2$, where we define $\hat{S}^{(i)}_x=\sum^{4}_{p=1}\hat{\sigma}_x^{(p,i)}$.
We also define
$\hat{S}^{(i)}_y=\sum^{4}_{p=1}\hat{\sigma}_y^{(p,i)}$
and $\hat{S}^{(i)}_z=\sum^{4}_{p=1}\hat{\sigma}_z^{(p,i)}$.
The parity operator has properties such as $\hat{P}_i\hat{S}^{(i)}_x\hat{P}_i=\hat{S}^{(i)}_x$, $\hat{P}_i\hat{S}^{(i)}_y\hat{P}_i=-\hat{S}^{(i)}_y$, and 
$\hat{P}_i\hat{S}^{(i)}_z\hat{P}_i=-\hat{S}^{(i)}_z$. In addition, the parity operator $\hat{P}_i$
commutes with $(\hat{S}_z^{(i)})^2$. 

The lowest eigenstate of $\hat{S}^{(i)}_x$ is described as $|S^{(i)}_x=-2\rangle $, and the second lowest eigenstate state is described as $|S^{(i)}_x=-1\rangle$.
We have $\hat{P}_i |S^{(i)}_x=-2\rangle = |S^{(i)}_x=-2\rangle$ and $\hat{P}_i |S^{(i)}_x=-1\rangle = -|S^{(i)}_x=-1\rangle$.
For the excited state search, we prepare the first excited state of $H^{(\rm{SVP})}_{\rm{D}}$ such as $|S^{(1)}_x=-1\rangle \bigotimes _{j=2}^N |S^{(j)}_x=-2\rangle$, and the parity of this state is $P_1=-1$ and $P_j=1$ for $j\geq 2$. 

For the SVP,
let us consider the case  $N=2$, $|\boldsymbol{b}_{1}|=\frac{1}{2}|\boldsymbol{b}_{2}|$, and $\boldsymbol{b}_{1}\cdot \boldsymbol{b}_{2}=0$. 
In this case, we have
$H^{(\rm{SVP})}_{\rm{D}}= \frac{B}{2}\hat{S}^{(1)}_x
+B\hat{S}^{(2)}_x$ and $H^{(\rm{SVP})}_{\rm{p}}=4(\hat{S}^{(1)}_z)^2 + (\hat{S}^{(2)}_z)^2$.
The first excited states of $H^{(\rm{SVP})}_{\rm{p}}$ are $|S^{(1)}_z=0\rangle |S^{(2)}_z=1\rangle$
and $|S^{(1)}_z=0\rangle |S^{(2)}_z=-1\rangle$.
Importantly, we have $\hat{P}_1|S^{(1)}_z=0\rangle=|S^{(1)}_z=0\rangle$. Therefore, the first excited states of $H^{(\rm{SVP})}_{\rm{p}}$ have the parity $\hat{P}_1=1$ while the first excited state of $H^{(\rm{SVP})}_{\rm{D}}$ has the parity $\hat{P}_1=-1$.
Since these state belong to different symmetry sectors, the excited state search by QA does not provide the solution of the SVP in this case \cite{imoto2022obtaining,francis2022determining}.

\bibliographystyle{apsrev4-1}
\bibliography{ura}
\end{document}